\documentclass[a4paper,twocolumn,amsmath,amssymb,aps,superscriptaddress,prd,nofootinbib, floatfix]{revtex4-2}  
\usepackage[utf8]{inputenc}

\usepackage{geometry}
\usepackage{graphicx}
\usepackage{mathtools}
\usepackage{siunitx}
\usepackage{xcolor}
\usepackage[normalem]{ulem}
\usepackage{url}
\usepackage{hyperref}
\usepackage{soul}
\usepackage{booktabs}
\usepackage{multirow}
\usepackage{commath}
\usepackage{comment}
\usepackage{xr}

\makeatletter
\newcommand*{\addFileDependency}[1]{
\typeout{(#1)}
\@addtofilelist{#1}
\IfFileExists{#1}{}{\typeout{No file #1.}}
}\makeatother

\newcommand*{\myexternaldocument}[1]{%
\externaldocument{#1}%
\addFileDependency{#1.tex}%
\addFileDependency{#1.aux}%
}

\myexternaldocument{suppl}

\begin{document}

\title{Wavelet--based tools to analyze, filter, and reconstruct transient gravitational--wave signals.}

\author{Andrea Virtuoso}
\email[Correspondence email address: ]{andrea.virtuoso@ts.infn.it}
\affiliation{Università di Trieste, Dipartimento di Fisica, I-34127 Trieste, Italy; \\ INFN Sezione di Trieste, I-34127 Trieste, Italy} 
\author{Edoardo Milotti}
\affiliation{Università di Trieste, Dipartimento di Fisica, I-34127 Trieste, Italy; \\ INFN Sezione di Trieste, I-34127 Trieste, Italy} 

\begin{abstract}
The analysis of gravitational--wave signals is one of the most challenging application areas of signal processing, because of the extreme weakness of these signals and of the great complexity of gravitational--wave detectors. Wavelet transforms are specially helpful in detecting and analyzing gravitational--wave transients and several analysis pipelines are based on these transforms, both continuous and discrete. While discrete wavelet transforms have distinct advantages in terms of computing efficiency, continuous wavelet transforms (CWT) produce smooth and visually stunning time--frequency maps where the wavelet energy is displayed in terms of time and frequency. In addition to wavelets, short--time Fourier transforms (STFT) and Stockwell transforms (ST) are also used, or the \textit{Q--transform}, which is a Morlet wavelet--like transform where the width of the Gaussian envelope is parameterized by a parameter denoted by $Q$ [Chatterji et al., Class. Quant. Gravity \textbf{21} (2004) S1809]. To date, the use of CWTs in gravitational--wave data analysis has been limited by the higher computational load when compared with discrete wavelets, and also by the lack of an inversion formula for wavelet families that do not satisfy the admissibility condition. In this paper we consider Morlet wavelets parameterized in the same way as the Q--transform (hence the name \textit{wavelet Q--transform}) which have all the advantages of the Morlet wavelets and where the wavelet transform can be inverted with a computationally efficient specialization of the non--standard inversion formula of Lebedeva and Postnikov [Lebedeva and Postnikov, Royal Society Open Science, \textbf{1} (2014) 140124]. We also introduce a two--parameter extension (the \textit{wavelet Qp--transform}) which is well--adapted to chirping signals like those originating from compact binary coalescences (CBC), and show that it is also invertible just like the wavelet Q--transform. The inversion formulas of both transforms allow for effective noise filtering and produce very clean reconstructions of gravitational--wave signals. Our preliminary results indicate that the method could be well suited to perform accurate tests of General Relativity by comparing modeled and unmodeled reconstructions of CBC gravitational--wave signals.
\end{abstract}

\maketitle


\section{Introduction}\label{sec:introduction}
Since the first detection of gravitational waves (GW) in 2015 \cite{abbott2016observation,abbott2016observing, abbott2016gw150914} the LIGO--Virgo--KAGRA (LVK) collaboration has successfully analyzed and reconstructed gravitational--wave transients (GWT) with different methods \cite{abbott2019gwtc,abbott2021gwtc,abbott2021gwtc3} which fall into two large classes, the \textit{modeled methods} and the \textit{unmodeled methods}. 
Modeled methods use theoretical waveforms and matched--filtering to detect and analyze GWTs \cite{veitch2015parameter,usman2016pycbc,dal2021real,aubin2021mbta,tsukada2023improved} and they have been hugely successful in the inference of the properties of merging compact binary objects (see, e.g., \cite{abbott2016gw150914}, or \cite{abbott2021gwtc3} and references therein).
Unmodeled methods are based instead on the coherence of data in different detectors, and make no assumptions on signal waveform \cite{abbott2017all, abbott2019all, abbott2021all, klimenko2016method,drago2021coherent,millhouse2018bayesian,cornish2021bayeswave, szczepanczyk2021observing, szczepanczyk2023search}.
Even though they are not as sensitive as modeled methods there are two main reasons to continue developing and using them: first, their non--reliance on prior models provides independent checks in the detection and analysis of GWTs \cite{abbott2016observing}. Second, unmodeled methods are effective also when dealing with GW sources where we lack precise theoretical waveforms, such as core--collapse supernovae (CCSN) \cite{abbott2020optically,szczepanczyk2021detecting,szczepanczyk2023optically}. 

\medskip 

The lack of well--defined reference waveforms requires the adoption of complex mathematical tools to extract signals from background noise.
The representations of signals in the time--frequency (TF) domain are specially useful to disentangle signals from noise and efficiently evaluate the coherence of data in different detectors. Such representations are obtained with a variety of methods, like short--time Fourier transforms (STFT), Stockwell transforms (ST) \cite{stockwell1996localization}, wavelet transforms (WT) \cite{mallat2008wavelet}, and several variants that mix the different approaches. 

While all transforms are superficially similar, they differ in important details, such as the existence of an inverse transform. 
The invertibility of a TF representation is crucial for further processing of the signals, such as filtering out unwanted noise. 
Continuous wavelet transforms (CWT) do have an exact reconstruction formula if and only if the mother wavelet $\psi(x)$ satisfies the \textit{admissibility condition}
\begin{equation}
    C_\psi = \int_{-\infty}^{+\infty} \frac{|\tilde{\psi}(y)|^2}{|y|} dy < \infty
\end{equation}
where $\tilde{\psi}(y)$ is the Fourier transform of $\psi(x)$. This is substantially equivalent to the condition that the mother wavelet has a vanishing integral  
\begin{equation}
\int_{-\infty}^{+\infty} \psi(x) dx = 0
\end{equation}
(see, e.g., \cite{daubechies1992ten}, Paragraph 2.4). When the admissibility condition is satisfied, the wavelet transform of the time--domain signal $x(t)$
\begin{equation}
    T_x(\nu,\tau) = \int_{-\infty}^{+\infty}x(t) \psi^*_{\nu,\tau}(t) dt
\end{equation}
is invertible with the standard formula 
\begin{equation}
\label{standardrec}
    x(t) = \frac{1}{C_\psi} \int_{-\infty}^{+\infty} \int_{-\infty}^{+\infty} d\nu\; d\tau\; T_x(\nu,\tau) \psi_{\nu,\tau}(t)
\end{equation}
Interestingly, other inversion formulas exist (see, e.g., \cite{daubechies1992ten}, Section 2.4) which also require the admissibility condition.

A TF representation should also have an excellent resolution, both in time and in frequency. However, time resolution $\sigma_{\tau}$ and frequency resolution $\sigma_{\nu}$ can not be both arbitrarily small, they are limited by the Gabor--Heisenberg principle \cite{gabor1946theory}
\begin{equation}
   \label{eq:Gabor_Heisenberg_generic}
   \sigma_{\tau}\sigma_{\nu} \geq \dfrac{1}{4\pi}
\end{equation}
The lower bound of this inequality is actually attained by the Morlet wavelets (see eqs. \eqref{eq:sigma_tau_Q} and \eqref{eq:sigma_phi_Q}).

Finally, the signal energy may be spread over regions of arbitrary shape in the TF plane, and this means that in order to achieve a sparse TF representation --- i.e., a representation that concentrates signal energy in as few components as possible -- we should be able to adapt the time and frequency resolutions to follow the main features of the TF representation of a signal. Sparse representations are specially useful in the analysis of GW signals, because sparseness favors signal detectability and helps signal filtering and reconstruction. 

Since the product of time resolution and frequency resolution is fixed, it should be possible to set a balance between resolutions to optimize the analysis of an input signal. As an example, STFT lacks this tunability because it has a fixed envelope window (and therefore  fixed time and frequency resolutions) for all frequencies: this implies that at different frequencies the window encompasses a different number of oscillations, which is often undesirable  \cite{rioul1991wavelets}. 
On the contrary, WTs feature envelope windows with variable width such that the number of oscillations is frequency--independent, resulting in variable time and frequency resolutions at different central frequencies. This scale invariance is a precious feature which we retain in our analysis of GWTs, just as in  past works \cite{necula2012transient,klimenko2016method,cornish2021bayeswave}, and which helps achieving sparse TF representations of signals. 

The issue of setting an optimal balance of time and frequency resolution is addressed by the Q--transform with a Gaussian window function \cite{brown1991calculation,chatterji2004multiresolution,chatterji2005search,robinet2020omicron, henshaw2024visualization}, which is defined by the pair of equations
\begin{subequations}
\label{origQtransf}
\begin{align}
   X(\tau,\nu,Q)&=\int_{-\infty}^{+\infty} dt \, s(t) g(t; \tau, \nu, Q)e^{-2\pi i \nu t}\label{eq:original_Q_transform}\\
    g(t; \tau, \nu, Q)&=\left(\frac{8\pi \nu^{2}}{ Q^{2}}\right)^{1/4} \, e^{-\left( \frac{2\pi \nu (t-\tau)}{Q} \right)^{2}}.
\end{align}
\end{subequations}
where $s(t)$ is the time--domain representation of the signal for which the Q--transform is evaluated, $\tau$ is the central time, $\nu$ the central frequency and $Q$ a parameter that sets the balance between time ($\sigma_{\tau}$) and frequency ($\sigma_{\nu}$) resolution.
The Q--transform is not a continuous wavelet transform: even if it inherits from wavelet transforms the property that its window $g(t; \tau, \nu, Q)$ depends on the frequency $\nu$ --- a key property of wavelets --- the oscillatory term $e^{-2\pi i \nu t}$ is not centered at $t=\tau$, as for wavelets, but is the same for all times.

Still, the Q--transform is very similar to the Morlet wavelet transform, and here we consider a version of the Morlet wavelets with the addition of the $Q$ parameter as in the Q--transform for tunability of the time and frequency resolutions. The wavelet version of the Q--transform that we consider in this paper drops the requirement of a common oscillatory term in the Q--transform, which is replaced by $e^{-2\pi i \nu (t-\tau)}$. 
With this substitution, we obtain a wavelet version of the Q--transform, the \emph{wavelet Q--transform}
\begin{equation}
\label{eq:wavelet_Q_transform}
    T(\tau,\nu,Q)=\int_{-\infty}^{+\infty} dt \, s(t) \psi^{*} (t; \tau, \nu)
\end{equation}
with the wavelets
\begin{equation}
    \psi^{*} (t; \tau, \nu, Q)=\left(\frac{8\pi \nu^{2}}{ Q^{2}}\right)^{1/4} \, e^{-\left( \frac{2\pi \nu (t-\tau)}{Q} \right)^{2}-2\pi i \nu (t-\tau)}.
       \label{eq:Gaussian_wavelet}
\end{equation}
where $^{*}$ denotes complex conjugate.
The time and frequency resolutions depend only on the Gaussian window, and are found to be the same for both Q--transform and wavelet Q--transform 
\begin{align}
\sigma_{\tau}^\mathrm{(Q)}&=\dfrac{Q}{4\pi\nu}
\label{eq:sigma_tau_Q} \\
\sigma_{\nu}^\mathrm{(Q)}&=\dfrac{\nu}{Q}
\label{eq:sigma_phi_Q}.
\end{align}
It is important to note that both resolutions depend on $\nu$. Moreover, while low values of $Q$ ($\sim 2\pi$) favor a better time resolution with respect to frequency resolution, high values of $Q$ ($\gg 2\pi$) give good frequency resolution and poor time resolution. 
In summary, when computing the wavelet Q--transform with high values of $Q$ the signal energy in the TF plane is more spread in time compared to low values of $Q$. In contrast, with low values of $Q$, the signal energy in the TF plane has a higher spread in frequency compared to high values of $Q$.
In all cases, the product of the resolutions remains fixed at the lower bound of Gabor--Heisenberg uncertainty inequality and endows this transform with the best possible TF resolution\footnote{In actual implementations the tails of the Gaussian window are truncated for two reasons: first, to improve computational efficiency, and second, because with real signals which are time-- and band--limited at some point this truncation has to be done anyway. As a result, the truncated wavelets have a slightly reduced resolution.}

Unfortunately, the Morlet wavelets do not match the admissibility condition, and the resulting WT is not invertible with the standard formula: since the wavelet Q--transform is just a parameterized version of the Morlet wavelet transform it is not invertible as well.\footnote{The BayesWave pipeline \cite{millhouse2018bayesian,cornish2021bayeswave}, which is based on Morlet wavelets, bypasses the non--invertibility issue by picking one by one individual wavelets within a Bayesian framework which utilizes a reversible--jump Markov chain Monte Carlo (RJMCMC), and then finally superposing all of them to reconstruct the signal. This method produces good reconstructions but the RJMCMC step is computationally very expensive. The non--invertibility of Morlet wavelets can also be overcome by shifting them to satisfy the admissibility condition \cite{addison2002low}: in GW data analysis, that possibility has been exploited in \cite{roy2022nonorthogonal}.}
However, recent progress on a wavelet inversion formula that does not require the admissibility condition \cite{lebedeva2014alternative,postnikov2016computational} means that the wavelet Q--transform can provide both TF representations comparable to those of the standard Q--transform and all the advantages of invertibility.
We make further progress in this direction and here we present a more efficient inversion formula which also implements TF filtering. This new inversion formula, which we discuss in Section \ref{sec:reconstruction_formula}, is a real game--changer, because it leads to computationally efficient denoising and waveform reconstruction with the wavelet Q--transform.

Next, in Section \ref{sec:Qp_generalities} we describe a variant of the wavelet Q--transform that is better adapted at describing GW chirps from CBCs and is still graced with an inverse formula. 
With this parameterization, GW chirps are better modeled and the Gaussian window can be made wider, as we discuss in Section \ref{subsec:transform_implementation}, giving in many cases a better match of the wavelet with the chirping signal, as  shown in Section \ref{sec:tests}.
Because of the new $p$ parameter, we call this new transform the \textit{wavelet Qp--transform}. 

In Section \ref{sec:tests} we apply the wavelet Qp--transform and the associated denoising method discussed in Section \ref{sec:reconstruction_formula}  to some well known GW events (GW150914, GW170817 and GW190521 \cite{abbott2016observation, abbott2016gw150914, abbott2016observing, abbott2017gw170817, abbott2020gw190521}), and compare the performance of the wavelet Qp--transform to that of wavelet Q--transform. 
We also evaluate the waveforms obtained with the denoising formula for both transforms, comparing these with a modeled reconstruction based on matched filtering.
Finally, we conclude the paper with a short discussion in Section \ref{sec:conclusions}.

\section{Non-standard reconstruction formula for the wavelet Q--transform}\label{sec:reconstruction_formula}

As explained in the Introduction, a major drawback of the Morlet wavelets, and therefore of the wavelet Q--transform, is that they do not satisfy the admissibility condition.
Here we introduce a modified form of the alternative wavelet reconstruction formula recently proposed by Lebedeva and Postnikov
\cite{lebedeva2014alternative,postnikov2016computational} which does not require the admissibility condition. A complete derivation of the reconstruction formula for the wavelet Q--transform is given in the Supplementary Text, and the result is
\begin{widetext}
\begin{equation}\label{eq:Q_transform_reconstruction_formula}
    s(\tau)=\dfrac{2}{\left(\dfrac{2}{\pi Q^{2}}\right)^{1/4}\text{Re}\left[\text{erf}\left(  \frac{Q}{2} \right)\right]} \, \text{Re}\left[\int_{0}^{+\infty} d\nu \, \dfrac{1}{i\sqrt{2\pi \nu}2\pi \nu}\dfrac{\partial}{\partial \tau}T(\tau,\nu,Q)\right].
\end{equation}
\end{widetext}
This new reconstruction formula is our first important result, which represents a considerable advantage with respect to the traditional Q--transform \cite{chatterji2004multiresolution,chatterji2005search,robinet2020omicron}. 
After discretization of the integral in eq. \eqref{eq:Q_transform_reconstruction_formula} as usually done with the standard formula \eqref{standardrec}   \cite{jordan1997implementation}, the new formula allows the application of specific TF filters.

\begin{figure}[ht!]
\centering
\includegraphics[width=.9\linewidth]{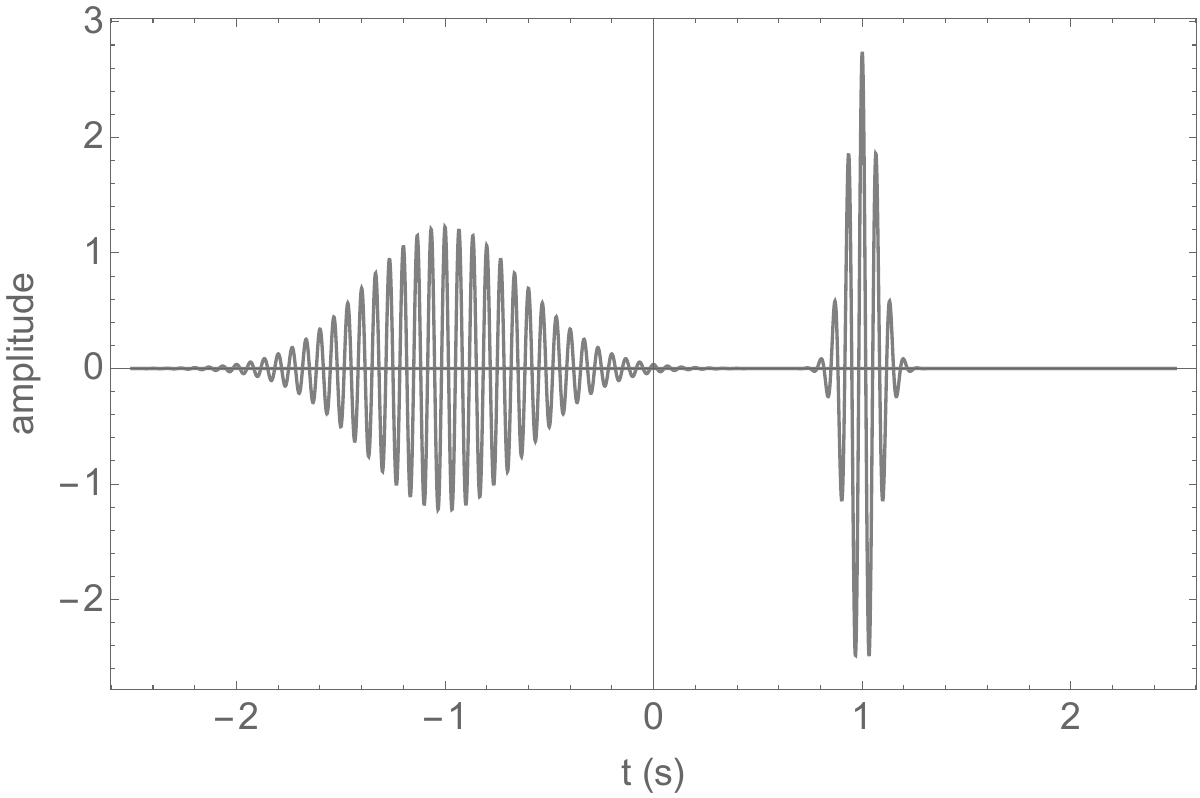}
\caption{Real part of a pair of Q--wavelets. The wavelet on the left has been generated with $Q=50$, $\tau=-1$~s, and $\nu=15$~Hz, the one on the right with $Q=10$, $\tau=1$~s, and $\nu=15$~Hz. The  standard deviation of the Gaussian envelope is $\approx 0.34$~s  for the wavelet on the left, and $\approx 0.07$~s for the wavelet on the right.}
\label{fig:morlet}
\end{figure}

\begin{figure}[ht!]
\centering
\includegraphics[width=.9\linewidth]{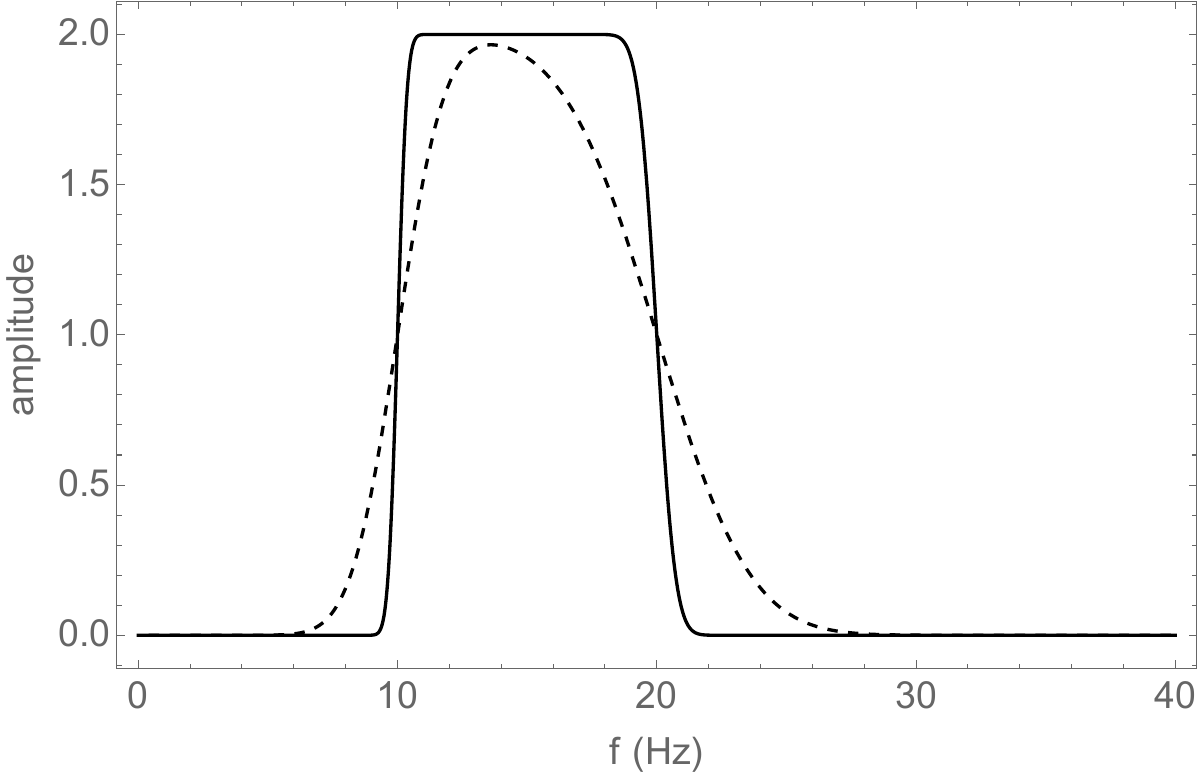}
\caption{Window functions for the pair of wavelets in Figure \ref{fig:morlet} and a single pair of values $\nu^\mathrm{low} = 10$~Hz and $\nu^\mathrm{high} = 20$~Hz. The solid line corresponds to $Q=50$ and the dashed line to $Q=10$. The wavelet with the longer duration produces a window with sharper edges in the frequency domain.}
\label{fig:filter}
\end{figure}

In general, we can perform filtering by multiplying the transform $T(\tau,\nu)$ times a filter function. When filtering means satisfying a predefined criterion C, e.g., passing an amplitude threshold, then the filter is represented by an indicator function $\mathbf{1}_C(\tau,\nu)$ equal to 1 wherever the criterion is satisfied and 0 elsewhere. We find that an important subclass of time--frequency filters, those that select only the regions bounded by the $L$ time--dependent frequency intervals\footnote{{We take into account  the possibility of having multiple unconnected intervals. This is important, e.g., when higher order modes are present in CBC chirps \cite{vedovato2022minimally}, or in the case of overlapping signals \cite{relton2022addressing}.}}  $\left[\nu^\mathrm{low}_{l}(\tau), \nu^\mathrm{high}_{l}(\tau)\right]$ with $l=1, \, ... \, , L$, can be implemented in a computationally efficient way with the following integral expression
\begin{equation}\label{eq:denoising_formula_Q}
    s_D(\tau)=\text{Re}\left[\int_{-\infty}^{+\infty} df \,\tilde{s}(f) e^{2\pi i f \tau} w(\tau,f,Q) \right],
\end{equation}
where $s_D(\tau)$ is the denoised signal and where we have introduced the TF window
\begin{widetext}
\begin{equation}\label{eq:denoising_window_Q}
    w(\tau,f,Q)=\dfrac{1}{\text{erf}\left(  \frac{Q}{2} \right)} \sum_{l} \left\{ \text{erf}\left[ \dfrac{Q}{2} \left( \dfrac{f-\nu^\mathrm{low}_{l}(\tau)}{\nu^\mathrm{low}_{l}(\tau)}\right) \right] - \text{erf}\left[ \dfrac{Q}{2} \left( \dfrac{f-\nu^\mathrm{high}_{l}(\tau)}{\nu^\mathrm{high}_{l}(\tau)}\right) \right] \right\}.
\end{equation}  
\end{widetext}
As an example, consider Figure \ref{fig:morlet} which shows the real part of a pair of Q-wavelets, and Figure \ref{fig:filter} which shows the corresponding window functions for a single pair of values $\nu^\mathrm{low}$ and $\nu^\mathrm{high}$. 

The window functions are equivalent to narrow bandpass filters in the frequency domain, with the larger $Q$'s producing a sharper filter functions. Neither filter has a compact support, however it is clear that the decay of the filter tails is very fast, and therefore both filters are very well localized in the sense first explained by Slepian \cite{slepian1976bandwidth,slepian1983some}. Notice also that both filter functions are real and therefore act as non--causal filters.

\medskip 

Just as the coherence among detectors, a careful selection of the frequency intervals in the TF plane can also be used to reject background noise; in the analyses that we report in this paper we retain only those TF intervals having an energy above a fixed threshold, as described in Section \ref{sec:tests}.

\section{The wavelet Qp--transform}\label{sec:Qp_generalities}

\subsection{A wavelet transform with chirping--frequency wavelets}

In this Section we introduce a variant of the wavelet Q--transform where we let the frequency of the wavelet change to adapt to the shape of gravitational wave chirps. 
A similar idea was developed previously, in the context of the so--called ``chirplets'' \cite{millhouse2018bayesian,mann1991chirplet,mann1995chirplet,candes2008detecting,chassande2006best,pai2008best,yu2016general}, with the goal of following the TF evolution of the CBC inspiral phase. This was achieved either by introducing the first derivative of the wavelet frequency as a parameter 
\begin{equation}
    \nu(t)=\nu(\tau)+\eval{\dfrac{\partial \nu}{\partial t}}_{t=\tau}(t-\tau).
\end{equation}
and by deploying a wide array of different methods to place chirplets in the TF plane \cite{millhouse2018bayesian,mann1991chirplet,mann1995chirplet,candes2008detecting,chassande2006best,pai2008best,yu2016general}, or by modeling the TF evolution of the signal with a more complex polynomial parameterization \cite{yang2012multicomponent,li2022chirplet}.
The wavelet Qp--transform is based on the same idea, but with an important difference: the derivative ${\partial \nu}/{\partial t}|_{t=\tau}$ is itself a function of the local wavelet frequency over the wavelet duration $[\tau-\sigma_{\tau}, \tau+\sigma_{\tau}]$:

\begin{align}
    \eval{\dfrac{\partial \nu}{\partial t}}_{t=\tau}&=\dfrac{(1+p)\nu(\tau)-(1-p)\nu(\tau)}{(\tau+\sigma_{\tau})-(\tau-\sigma_{\tau})}=4\pi\nu^{2}(\tau)\dfrac{p}{Q}\nonumber\\
    &=\dfrac{1}{2\pi}\left( \dfrac{2\pi\nu(\tau)}{Q}\right)^{2}2pQ,
    \label{eq:frequency_derivative}
\end{align}  
where we have used expression
\eqref{eq:sigma_tau_Q} for $\sigma_{\tau}$. The $p$ parameter determines the fractional frequency change with respect to the central frequency over the $[\tau-\sigma_{\tau}, \tau+\sigma_{\tau}]$ time interval. The resulting nonlinear frequency change helps these wavelets in adapting to the CBC chirping gravitational--wave signals. 

As a result, the (conjugate) Qp--wavelets are described by the following expression:
\begin{widetext}
\begin{equation}
\label{Qpwavelet}
    \psi^{*} (t; \tau, \nu, Q, p)=\left(\frac{8\pi \nu^{2}}{ Q^{2}}\right)^{1/4} \, e^{-\left( 2\pi \nu (t-\tau)\frac{\sqrt{1+2ipQ}}{Q} \right)^{2}-2\pi i \nu (t-\tau)}.
\end{equation}
\end{widetext}

\begin{figure}[ht!]
\centering
\includegraphics[width=.9\linewidth]{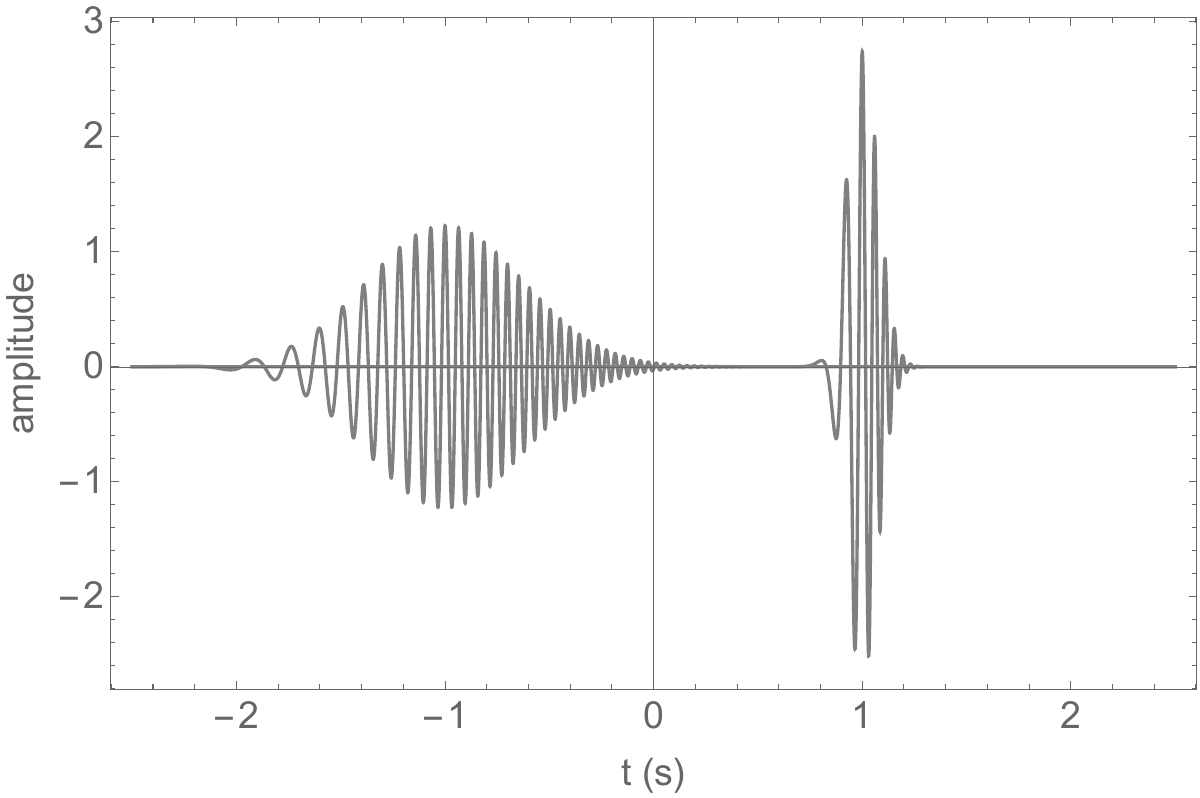}
\caption{Real part of a pair of Qp--wavelets. The wavelet on the left has been generated with $Q=50$, $\tau=-1$~s, $\nu=15$~Hz, and $p=0.1$, the one on the right with $Q=10$, $\tau=1$~s, $\nu=15$~Hz, and $p=0.1$. Because of the positive value of $p$ the frequency increases in the positive direction of time.}
\label{fig:Qp-wavelets}
\end{figure}

Figure \ref{fig:Qp-wavelets} shows a pair of Qp--wavelets with different central frequencies and positive $p$, which determines a chirp with frequency increasing in the positive directions, just as the CBC gravitational--wave chirps.

It is important to note that for Qp--wavelets the time resolution is the same as in eq. \eqref{eq:sigma_tau_Q}, while the frequency resolution differs from that found in eq. \eqref{eq:sigma_phi_Q}. Indeed, it can be shown that in the case of the Qp wavelets 

\begin{align}
    \label{eq:timeres}
    \sigma_{\tau}^\mathrm{(Qp)}&=\dfrac{Q}{4\pi\nu}\\
    \label{eq:freqres}
    \sigma_{\nu}^\mathrm{(Qp)}&=\dfrac{\nu}{Q}\sqrt{1+(2pQ)^{2}}.
\end{align}
\noindent
so that the lower bound of the Gabor--Heisenberg uncertainty product is no longer reached if $p\neq 0$. 
This reduced resolution comes from the frequency spread due to the chirping behavior and is discussed further in Section \ref{subsec:transform_implementation}.
Using the Qp wavelets in eq. \eqref{Qpwavelet} we can write the wavelet Qp--transform of a signal $s(t)$ as

\begin{widetext}
\begin{subequations}
\label{waveletQptransf}
\begin{align}
\label{eq:Qp_transform_time_representation}
    T(\tau,\nu,Q,p)&=\int_{-\infty}^{+\infty} dt \, s(t) \psi^{*} (t; \tau, \nu, Q, p) \\
    &=\int_{-\infty}^{+\infty} dt \, s(t) \left(\frac{8\pi \nu^{2}}{ Q^{2}}\right)^{1/4} \, e^{-\left( 2\pi \nu (t-\tau)\frac{\sqrt{1+2ipQ}}{Q} \right)^{2}}e^{-2\pi i \nu (t-\tau)} \\
\label{eq:Qp_transform_frequency_representation}
  &= \int_{-\infty}^{+\infty} df \,\tilde{s}(f)\tilde{\psi}^{*} (f; \tau, \nu, Q, p) \\
  &= \int_{-\infty}^{+\infty} df \,\tilde{s}(f) \left(\frac{1}{2\pi \nu^{2} Q^{2}}\right)^{1/4} \frac{Q}{\sqrt{1+2iQp}} e^{-\left( \frac{Q}{2\sqrt{1+2iQp}} \frac{f-\nu}{\nu} \right)^{2}}e^{2\pi i f \tau} 
\end{align}
\end{subequations}
\end{widetext}

where Fourier transforms are denoted by the overhead tilde.
When computing the wavelet Qp--transform, the $p$ parameter aligns the wavelets along the chirp: positive values of $p$ are associated with a counterclockwise rotation, which results in a better match of the transform with frequency--increasing signals; negative values of $p$ are associated with a clockwise rotation, which results in a better match of the transform with frequency--decreasing signals. Therefore, the reduced frequency resolution is offset by the better match of wavelet and signal.
If $p=0$, i.e., if there is no frequency chirp, we recover all the results obtained in the case of the wavelet Q--transform.

\subsection{Non-standard inversion formula for wavelet Qp--transform}\label{subsec:Qp_inversion_formula}

The inversion formula for the wavelet Q--transform, eq. \eqref{eq:Q_transform_reconstruction_formula}, as well as its denoising version,  eqs. \eqref{eq:denoising_formula_Q} and \eqref{eq:denoising_window_Q}, can be extended to the wavelet Qp--transform. 
The resulting formulas, which are proved in the Supplementary Text, are:
\begin{widetext}
\begin{equation}\label{eq:Qp_transform_reconstruction_formula}
    s(\tau)=\dfrac{2}{\left(\dfrac{2}{\pi Q^{2}}\right)^{1/4}\text{Re}\left[\text{erf}\left(  \frac{Q}{2\sqrt{1+2iQp}} \right)\right]} \, \text{Re}\left[\int_{0}^{+\infty} d\nu \, \dfrac{1}{i\sqrt{2\pi \nu}2\pi \nu}\dfrac{\partial}{\partial \tau}T(\tau,\nu,Q,p)\right].
\end{equation}
\begin{equation}\label{eq:denoising_formula_Qp}
    s_D(\tau)=\text{Re}\left[\int_{-\infty}^{+\infty} df \,\tilde{s}(f) e^{2\pi i f \tau} w(\tau,f,Q,p) \right]
\end{equation}
\begin{equation}\label{eq:denoising_window_Qp}
    w(\tau,f,Q,p)=\dfrac{1}{\text{Re}\left[\text{erf}\left(  \frac{Q}{2\sqrt{1+2iQp}} \right)\right]} \sum_{l} \left[ \text{erf}\left( \dfrac{Q}{2\sqrt{1+2iQp}} \left( \dfrac{f-\nu^\mathrm{low}_{l}(\tau)}{\nu^\mathrm{low}_{l}(\tau)}\right) \right) - \text{erf}\left( \dfrac{Q}{2\sqrt{1+2iQp}} \left( \dfrac{f-\nu^\mathrm{high}_{l}(\tau)}{\nu^\mathrm{high}_{l}(\tau)}\right) \right) \right].
\end{equation}
\end{widetext}
Again, notice that setting $p=0$ we recover the previous reconstruction formulas for the wavelet Q--transform. It is important to remark that they critically depend on the parameterization of the frequency derivative, which leads to a solvable Gaussian integral.

\begin{figure}[ht!]
\centering
\includegraphics[width=0.9\linewidth]{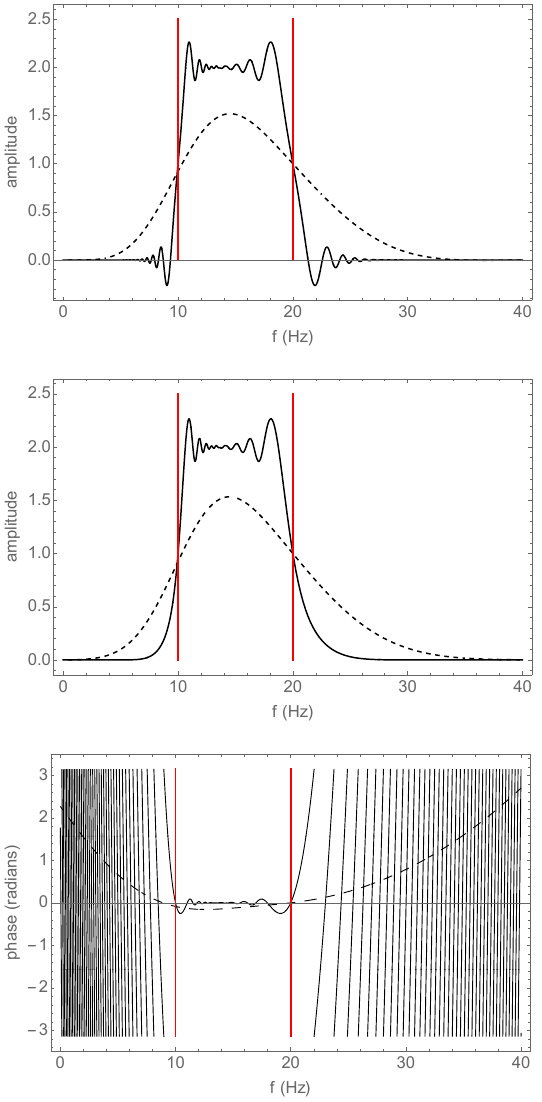}
\caption{Window function determined by Qp--wavelets with different $Q$'s. Here we take a single pair of values $\nu^\mathrm{low} = 10$~Hz and $\nu^\mathrm{high} = 20$~Hz (marked by the red vertical lines), $\nu=15$~Hz, and $p=0.05$. In each panel, the black solid line corresponds to $Q=100$, and the black dashed line corresponds to $Q=5$. The top panel shows the real part of the window function $w(\tau,f,Q,p)$, the middle panel shows its absolute value, the bottom panel shows its phase (in radians). Outside the 10~Hz -- 20~Hz interval, the phase of the window function oscillates wildly. }
\label{fig:Qp-filter}
\end{figure}

Figure \ref{fig:Qp-filter} shows an example of the complex window $w(\tau,f,Q,p)$ in the case of the wavelet Qp--transform, which is a complex function.
We notice that the window amplitude in the frequency domain (middle panel) is slightly irregular: this is a manifestation of the Gibbs phenomenon for this particular kind of wavelets. It is also important to note that the window phase is frequency--dependent (bottom panel): this window acts again as a filter, but, unlike the window for the wavelet Q--transform, it does not behave as a non--causal filter as a consequence of the asymmetrical shape of the wavelet in the time domain. Still, the phase is close to zero between $\nu^\mathrm{low}$ and $\nu^\mathrm{high}$, and this means that for frequencies in this range the window behavior is close to that of a non--causal filter. Numerical experiments indicate that, as expected, the non--causal behavior is more prominent for larger values of $p$ and for higher ratios $(\nu^\mathrm{high}-\nu^\mathrm{low})/Q$. This means that by keeping these values low, in specific applications we can safely use the phase information provided by the wavelet Qp--transform to complement the amplitude information.

\subsection{Practical implementation of the wavelet Q-- / Qp--transform}\label{subsec:transform_implementation}

The continuous wavelet transforms that we consider in this paper are applied to data samples $s(t_{n})$ ($n=0,...,N-1$) taken with sampling rate $f_{s}$. The samples are whitened using an interpolated version of the power spectral density that defines the spectral sensitivity of the detector and which is independent from the sample rate with which the power spectral density is obtained. On the other hand, since the width of the frequency bins is a function of the sampling rate each frequency bin in the transform of a sampled signal picks up a total noise power which is a function of the sampling rate. For this reason, a correct comparison with the noise background, and therefore a correct whitening, requires a transform that does not depend on the sampling rate. In the Supplementary Text we prove that the following {scaled (dimensionless)} version of equation \eqref{eq:Qp_transform_frequency_representation} satisfies this requirement
\begin{equation}
    T_\mathrm{nd}(\tau,\nu,Q,p) = \dfrac{\sqrt{f_{s}}}{N}\sum_{m}\,\tilde{s}(f_{m})\tilde{\psi}^{*} (f_{m}; \tau, \nu, Q, p)\label{eq:frequency_domain_Qp_implementation}
\end{equation}
where $\tilde{s}(f_{m})=\sum_{n}s(t_{n})e^{-2\pi i f_{m}t_{n}}$ and $m=-N/2+1, ..., +N/2$.
As a result, the individual values of the  wavelet Qp--transform  $|T(\tau,\nu, Q,p)|^2$ corresponding to the whitened noise background, i.e., with Gaussian white noise $~N(0,1)$,  follow a $\chi^{2}$ distribution with $2$ degrees of freedom, with mean value and standard deviation equal to $2$ for all reasonable values of $Q$ and $p$ involved in GW data analysis (see the Supplementary Text for details).

In addition to the obvious discretization of time $t$ and frequency $f$ due to sampling, and to the numerical problems that this sampling carries with it \cite{jordan1997implementation}, we must also select a discrete tiling of the TF plane $(\tau,\nu)$ to estimate the continuous wavelet transforms in a graphically satisfactory and computationally efficient way. To this end, we address the following two issues:
\begin{itemize}
    \item choice of a tiling represented by a discrete lattice of points $(\tau_i, \nu_j)$ in the  TF plane  for the evaluation and representation of the transform, assuming a time range $[\tau_\mathrm{min}, \tau_\mathrm{max}]$ and a frequency range $[\nu_\mathrm{min}, \nu_\mathrm{max}]$;
    \item informed selection of $Q$ and $p$ to produce a signal representation as sparse as possibile.    
\end{itemize}

We can link the positions $(\tau_i, \nu_j)$ of the tiles to the TF resolutions, eq. \eqref{eq:timeres} and eq. \eqref{eq:freqres}, of the wavelets in the following way: first, we consider a set of frequencies $\{\nu_{j}\}$ covering the $[\nu_\mathrm{min}, \nu_\mathrm{max}]$ range such that
\begin{equation}    \nu_{j+1}=\nu_{j}+\alpha\sigma_{\nu_j}=\nu_{j}\left(1+\frac{\alpha}{Q}\sqrt{1+(2pQ)^{2}}\right)
\end{equation}
where the tiling parameter $\alpha$ has been introduced and where the notation $\sigma_{\nu_j}$ stresses the actual dependence of the uncertainty on $\nu_{j}$.
Next, for each frequency $\nu_{j}$, the times $\tau_{i}(\nu_{j})$ are chosen accordingly
\begin{equation}
    \tau_{i+1}(\nu_{j})=\tau_{i}(\nu_{j})+\alpha \sigma_{\tau_{i}}(\nu_{j})=\tau_{i}(\nu_{j})+\alpha\dfrac{Q}{4\pi \nu_{j}},
\end{equation}
where, again, the notation stresses the dependence on frequency. The resulting tiling is similar to that used for the multiresolution frames of discrete wavelets \cite{daubechies1992ten}. Since the tile area is the constant $\alpha^2 \sqrt{1+(2pQ)^{2}}/4\pi$ and the time and frequency ranges correspond to the TF area 
\[(\tau_\mathrm{max}-\tau_\mathrm{min})(\nu_\mathrm{max}-\nu_\mathrm{min})\]
the total number of tiles is 
\begin{equation}
    N_\mathrm{tiles}=\dfrac{(\tau_\mathrm{max}-\tau_\mathrm{min})(\nu_\mathrm{max}-\nu_\mathrm{min})}{\alpha^{2}}\dfrac{4\pi}{\sqrt{1+(2pQ)^{2}}}.
\end{equation}
Notice that a smooth representation of the continuous wavelet transform requires a very redundant representation where $\alpha\le1$, and that the computational load for the wavelet Q-- and Qp--transforms scales as 
\[ \dfrac{(\tau_\mathrm{max}-\tau_\mathrm{min})(\nu_\mathrm{max}-\nu_\mathrm{min})}{\alpha^{2}}.\]

\bigskip 

Moving on to the selection of optimal values of $Q$ and $p$, we note first that we certainly need at least a few large--amplitude cycles in the wavelet for a good determination of the instantaneous frequency of a signal. We can compute the number of cycles $k$ in the range $[\tau-n\sigma_{\tau},\tau+n\sigma_{\tau}]$ by integrating the phase $2\pi \nu(t)$ over $t$ in the same range and dividing by $2 \pi$, i.e., 
\begin{equation}
  k=\int_{\tau-n\sigma_{\tau}}^{\tau+n\sigma_{\tau}} \left[ \nu(\tau) + \eval{\dfrac{\partial \nu}{\partial t}}_{t=\tau} (t-\tau)\right] \, dt.
\end{equation}
Using eq. \eqref{eq:frequency_derivative}, with little additional algebra we find
\begin{equation}
    Q=\dfrac{k4\pi}{2n+n^{2}p}.
\end{equation}
Setting $k>1$ and assuming $p\ll 1$, we find the bound 
\begin{equation}
    Q>\dfrac{4\pi}{2+p} \sim 2\pi.
\end{equation}
Therefore, from now on we take $Q\gtrsim2\pi$.

\medskip 

There are no similar bounds on $p$ although we expect a limited frequency evolution within the same $[\tau-n\sigma_{\tau},\tau+n\sigma_{\tau}]$ range, which leads to the rule--of--thumb $p \lesssim 1/Q$.

\medskip

Since we aim at a sparse representation \cite{mallat2008wavelet} after removing the noise background, we filter out the tiles in the TF plane that are below a preset energy threshold and select those values of $Q$ and $p$ that maximise the energy density $\varepsilon$ of the surviving tiles
\begin{equation}\label{eq:energy_density_definition}
\varepsilon=\dfrac{\int\int_{|T(\tau,\nu,Q,p)|^{2}>\mathrm{thr}}|T(\tau,\nu,Q,p)|^{2}d\tau d\nu}{\int\int_{|T(\tau,\nu,Q,p)|^{2}>\mathrm{thr}}d\tau d\nu}.
\end{equation}
The procedure satisfies two goals: it maximizes the total energy of the selected tiles and at the same time it minimizes their TF area, leading to a representation of the signal as compact as possible.

In this context, it is interesting to compare the wavelet Q--transform and the wavelet Qp--transform: since Qp wavelets better adapt to a chirp, they can have a longer duration, i.e., a larger $Q$, and fewer above--threshold wavelets are actually needed to represent a chirp. This gain in terms of sparsity of wavelet representation is balanced by a loss in time resolution: a larger $Q$ corresponds to a larger TF uncertainty product.

\section{Application to a few notable GW events}\label{sec:tests}

In this Section we apply the wavelet Q-- and Qp--transforms to three well--known events detected during the first three LIGO--Virgo observing runs: GW150914, GW170817 and GW190521.
The data are public and are taken from the Gravitational--Wave Open Science Center website (GWOSC) \cite{abbott2021open,abbott2023open}. In each case, data sampled at $\SI{4096}{Hz}$ are downloaded, downsampled at $\SI{2048}{Hz}$, and finally whitened before analysis \footnote{For the decimation step we used the SciPy \cite{2020SciPy-NMeth} function \texttt{scipy.signal.decimate}, and for the whitening step the GWpy \cite{macleod2021gwpy} method \texttt{whiten} defined in the class \texttt{TimeSeries}.}.

\subsection{GW150914}

The first GW event was detected by the LIGO--Virgo collaboration in 2015 \cite{abbott2016observation} and widely analyzed with both match--filtered methods and with unmodeled methods \cite{abbott2016gw150914, abbott2016observing}. Here, we reanalyze the public data using both the wavelet Qp-- and the wavelet Q--transform on LIGO Livingston (L1) and LIGO Hanford (H1) detectors' data, evaluating the maximum of the energy (``energy peak''), the energy density above a given energy threshold defined in eq. \eqref{eq:energy_density_definition} and the corresponding TF area.
It is worth mentioning that even if obtained with the wavelet Q--transform our TF map looks very similar to those made with the conventional Q--transform because in these representations the only difference is a phase term that disappears when evaluating the local signal energy $|T_\mathrm{nd}(\tau,\nu,Q,p)|^{2}$.

\begin{figure}[ht!]
\centering
\includegraphics[width=\linewidth]{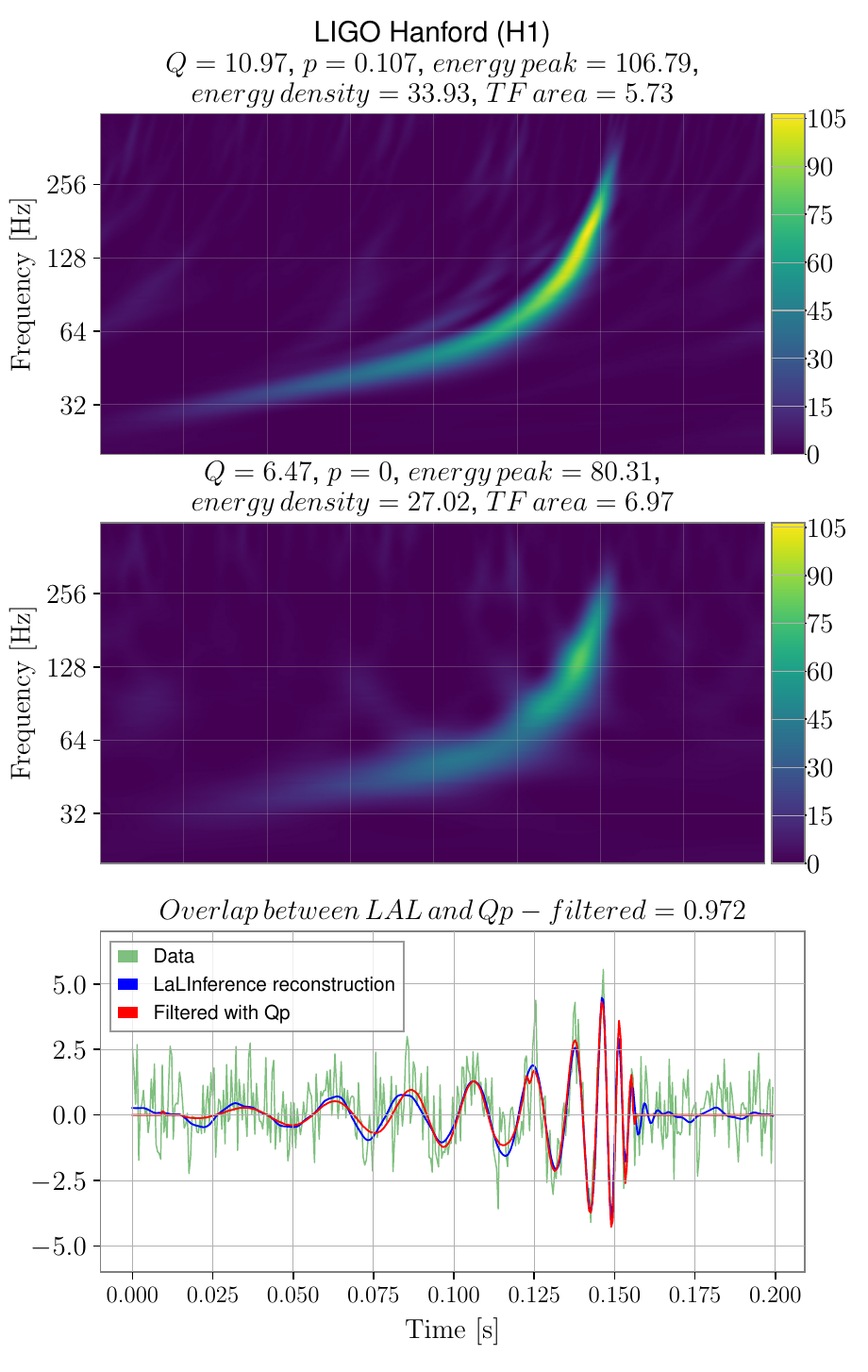}
\caption{Q-- and Qp--reconstruction of the first event GW150914, using data from H1. The top panel shows the  wavelet Qp--transform and the middle panel shows the wavelet Q--transform; the colorbars display the energy scale, i.e., the value of $|T_\mathrm{nd}(\tau,\nu,Q,p)|^{2}$. The bottom panel shows the strain plot with the original data (green), the waveform reconstructed with LALInference (blue) (from \cite{abbott2016observation}), and the waveform from the denoising formula for the wavelet Qp--transform (red), eqs. \eqref{eq:denoising_formula_Qp} and \eqref{eq:denoising_window_Qp}.}
\label{fig:GW150914}
\end{figure}

We estimate the GW waveforms using the denoising formula for both wavelet transforms, eqs. \eqref{eq:denoising_formula_Qp}, \eqref{eq:denoising_window_Qp}, and eqs. \eqref{eq:denoising_formula_Q}, \eqref{eq:denoising_window_Q}.
In each case we evaluate the residuals between data and the reconstructed waveforms: a good reconstruction is expected to produce a Gaussian distribution of the residuals, due to the whitening performed before the analysis. The resulting TF maps and the reconstructed waveforms for the Hanford interferometer are shown in Figure \ref{fig:GW150914}. The Supplementary Text includes an extended version of this Figure and a summary table with evaluations of the performance of the transforms. 

\medskip

The wavelet Qp--transform produces a considerable increase in both peak energy and energy density with respect to the wavelet Q--transform: the increase of the energy density is partly due to the corresponding lowering of the TF--area, meaning that the wavelet Qp--transform produces a more compact TF representation of the signal, as expected.
Note that these values depend on the threshold chosen when calculating the energy density.
Here, the energy threshold has been set to $E_\mathrm{thr}=7$: indeed, for a Gaussian noise background, we expect only $0.1\%$ of the transform values to be above that threshold, as shown in the Supplementary Text.

The waveforms (bottom panel in Fig. \ref{fig:GW150914}) have been obtained applying the denoising formula to the corresponding wavelet transform: in each case, pixels have been selected  if $|T_\mathrm{nd}(\tau,\nu,Q,p)|^{2}>E_\mathrm{thr}=7$.
The signal shapes obtained with the wavelet Q-- and  Qp--transforms are comparable to those obtained with LALInference, a pipeline based on matched--filtering and Bayesian estimation \cite{veitch2015parameter}, and reported in \cite{abbott2016observation, abbott2016gw150914}. 
We evaluate the similarity between the waveforms reconstructed with the wavelet Q-- and  Qp--transforms and the LAL reconstruction with the \textit{overlap}:
\begin{equation}
O\left[s_\mathrm{Qp/Q},s_\mathrm{LAL}\right]=\dfrac{(s_\mathrm{Qp/Q}, s_\mathrm{LAL})}{\sqrt{(s_\mathrm{Qp/Q}, s_\mathrm{Qp/Q}) \cdot(s_\mathrm{LAL}, s_\mathrm{LAL})}},
\end{equation}
where the subscripts denote the kind of reconstruction and $(a,b)=\int_{t_{rs}} a(t)b(t)dt$ is the scalar product calculated over the time duration of the reconstructed signal.
Both transforms return an overlap very close to $1$ (the ideal match), as reported in Table 1 in the Supplementary Text: this is a remarkable result, considering that we used neither templates as in modeled analyses \cite{veitch2015parameter} nor the likelihood formalism of unmodeled analysis \cite{klimenko2016method}.
The effectiveness of the denoising procedure is highlighted by the histograms of the residuals, which have a distribution close to Gaussian for both transforms and is corroborated by the values of the standard deviations reported in Table 1 in the Supplementary Text.
Overall, there is no major difference between the waveforms obtained from wavelet Qp-- and wavelet Q--transform: this is probably due to the fact that the signal energy is sufficiently high to be mostly above threshold for both transforms. This is not the case for the Virgo data of GW190521, which we discuss in the next subsection.

\subsection{GW190521}

\begin{figure}[ht!]
\centering
\includegraphics[width=\linewidth]{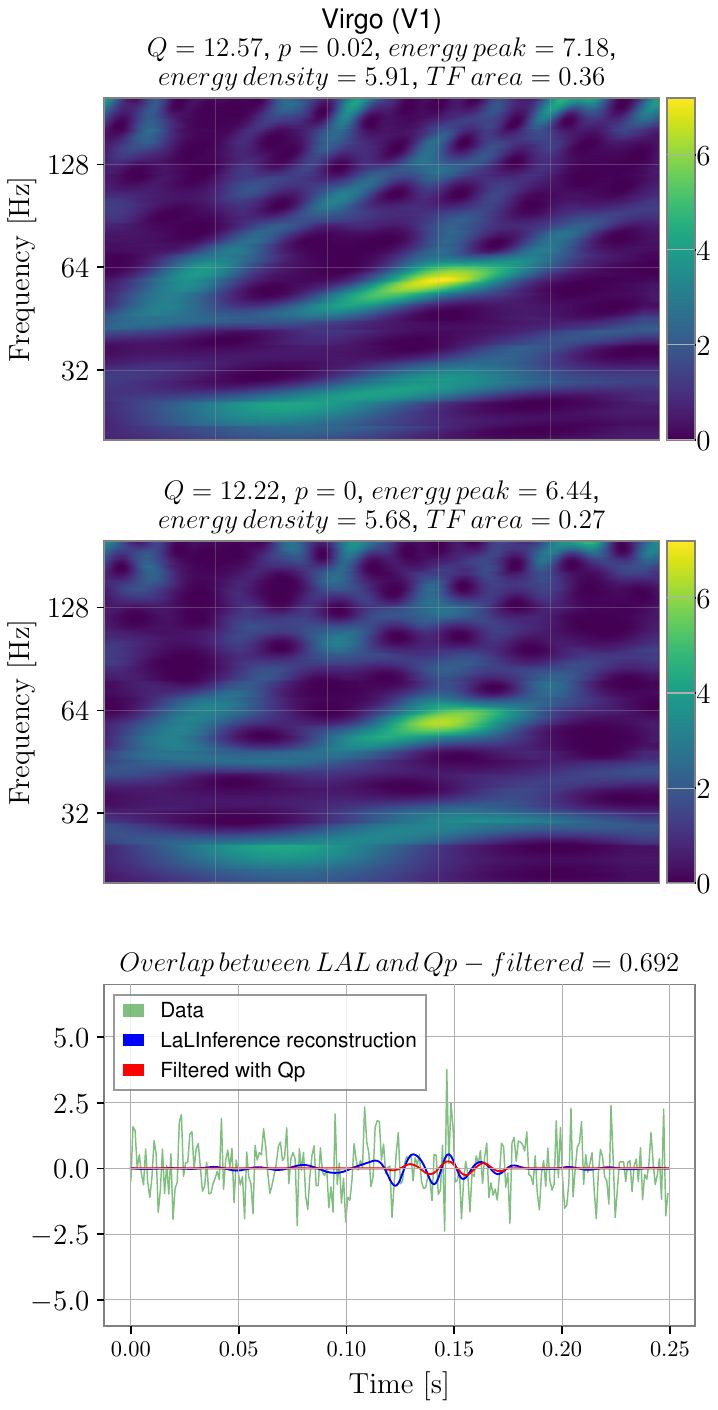}
\caption{Q-- and Qp--reconstruction of GW190521, using data from V1. The top panel shows the  wavelet Qp--transform and the middle panel shows the wavelet Q--transform; the colorbars display the energy scale, i.e., the value of $|T_\mathrm{nd}(\tau,\nu,Q,p)|^{2}$. The bottom panel shows the strain plot with the original data (green), the waveform reconstructed with LALInference (blue) (from \cite{abbott2020gw190521}), and the waveform from the denoising formula for the wavelet Qp--transform (red), eqs. \eqref{eq:denoising_formula_Qp} and \eqref{eq:denoising_window_Qp}.}
\label{fig:GW190521}
\end{figure}

The analysis for GW190521 \cite{abbott2020gw190521} is performed in the same way as done for GW150914, with the addition of Virgo (V1) data. 
In this case, Figure \ref{fig:GW190521} shows the TF map and the reconstructed waveform obtained with the wavelet Qp--transform and V1 data; an extended version of the Figure can be found in the Supplementary Text, which also includes a quantitative summary of the results in Table 2.

This event differs from the previous one in many ways, it has a lower signal--to--noise ratio, its duration is shorter and the bandwidth is limited to lower frequencies, and it has been detected by a three--interferometer network. V1 is the least sensitive interferometer, and for this reason we lowered the denoising threshold to $E_\mathrm{thr}=5$, mainly to avoid cutting too large a portion of the signal. Still, this threshold is  sufficiently high to effectively remove Gaussian noise, since in that case only $0.7\%$ of the transform values surpass this energy threshold.
As a general rule, the choice of the energy threshold can be made by evaluating the energy distribution of the transform values, identifying the signal by distinguishing it from the $\chi^{2}$ distribution of the Gaussian noise, and finally choosing the threshold by setting a reasonable trade--off between the removal of Gaussian noise and the preservation of the GW signal.

The complete results are shown in Figure 3 in the Supplementary Text, where
the thresholds for L1 and H1 have been kept at $E_\mathrm{thr}=7$ as for GW150914. In general, we note that the wavelet Qp--transform always performs better than the wavelet Q--transform, as reported in Table 2 in the Supplementary Text.
For Virgo the improvement of the  wavelet Qp--transform with respect to the wavelet Q--transform is more marked and it is important to note that results are considerably improved by careful filtering, even though the threshold has been lowered with respect to that used for the other detectors. With the wavelet Qp--transform several pixels pick up a higher energy density and cross the energy threshold. Overall, the reduction of the TF area due to the compactness of the wavelet Qp--transform representation is overcompensated by the larger number of above--threshold pixels, leading to a noticeable increase of the TF area of the wavelet Qp--transform with respect to the wavelet Q--transform.
In other words, with the wavelet Q--transform many pixels that are likely to originate from the signal remain below threshold, while with the wavelet Qp--transform they move above threshold, a clear  improvement due to the introduction of the $p$ parameter, and overall, the energy density also increases. We remark that the wavelets of both transforms are energy--normalized, therefore any difference in energy is given only by the better/worse match of the signal with the wavelets.

\medskip 

The waveforms obtained with the denoising formulas applied to the TF map obtained with both wavelet transforms are quite good for L1 and H1 and comparable with those obtained with LALInference and reported in \cite{abbott2020gw190521}, with a high value of the overlap.
The low energy values for both transforms affect the Virgo reconstruction, and in that case the overlap with LAL reconstruction is lower.
However, the wavelet Qp--transform has a larger number of above--threshold pixels and a much better waveform reconstruction, resulting in a considerably higher overlap with the LAL reconstruction with respect to that of Q--transform.

\subsection{GW170817 and glitch denoising}

A peculiarity of the GW170817 detection \cite{abbott2017gw170817} is that about a second before the merger a loud glitch was observed in the L1 data: in the analysis performed by the LIGO--Virgo collaboration the glitch was successfully removed with noise mitigation techniques \cite{abbott2016gw150914, usman2016pycbc, abbott2018effects}.
Here, we focus on the removal of this glitch using the Qp--wavelet denoising formula.
We do so by performing first a wavelet Qp--transform of the original data by using the combination of $Q$ and $p$ values giving the best representation of the GW signal: then, we apply the denoising formula setting to zero all pixels above an energy threshold of $25$. 
Again, this threshold has been chosen by evaluating the transform energy distribution, identifying both the signal and the glitch distributions, and finally taking a threshold that does not cut off the signal while removing the glitch only.
In this way, we obtain a denoised TF map and a filtered time series, for which we compute again the wavelet Qp--transform with the same $Q$ and $p$ pair maximizing the energy density of the GW signal.

\begin{figure}[ht!]
\centering
\includegraphics[width=\linewidth]{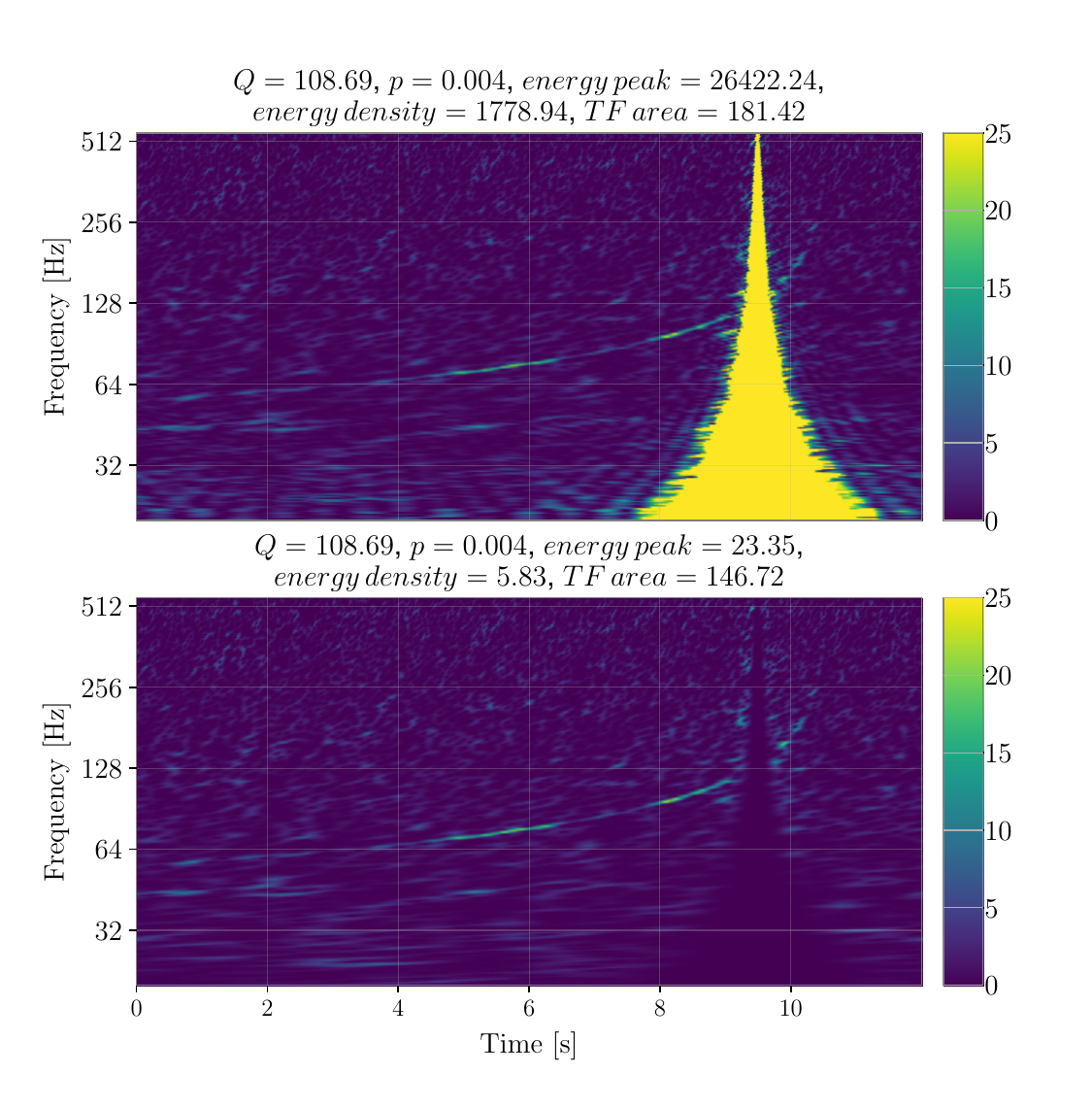}
\caption{Wavelet Qp--transform of the L1 data for GW170817. The upper panel shows the wavelet Qp--transform of the original whitened data which include the glitch. The lower panel shows the filtered TF map, where the glitch has been almost completely removed.}
\label{fig:GW170817}
\end{figure}

Our results are shown in the two panels of Figure \ref{fig:GW170817}: the wavelet Qp--transform of the filtered series (bottom panel) shows that the glitch (well visible in the upper panel) has been almost completely removed.

\section{Conclusions}\label{sec:conclusions}

In this paper we presented important improvements to the Q--transform.
We extended the Q--transform to the wavelet Q--transform --- a parameterized version of the Morlet wavelet transform --- which does have an inversion formula, unlike the Q--transform. This result extends the applicability of the formula originally obtained by Lebedeva and Postnikov \cite{lebedeva2014alternative}, leading to an original denoising algorithm which is both effective and computationally efficient. 
While the wavelet Q--transform, just as the Q--transform, has a minimal uncertainty in the time-frequency plane, it does not have an optimal performance in the case of chirping signals like those produced by CBCs. We found a useful variant --- which we dubbed the \textit{wavelet Qp--transform} --- which can still be inverted and produces sparser representations of the gravitational--wave signals from CBCs. We have fully characterized these wavelets, both in the time and in the frequency domain, also studying their statistical fluctuations and correlation properties in presence of a Gaussian white noise background. We have produced a Python implementation of both transforms and used it to analyze some important gravitational--wave signals detected by the LIGO/Virgo Collaboration during the O1 and O2 observing runs. These tests illustrate the higher efficiency of the wavelet Qp--transform for chirping signals, and also the excellent performance of the transform as a denoising tool. 

\medskip

The results presented here are encouraging, even though we have tested the transforms on a very small set of GW events, and they suggest more systematic studies involving a larger number of events or a set of simulations. 

There remain several unanswered questions. The most pressing is how to combine the data streams of several gravitational--wave detectors exploiting their coherence. 

\bigskip

Finally, we wish to remark that the inversion formula described  in this paper, in particular its denoising version of eqs. \eqref{eq:denoising_formula_Q} and \eqref{eq:denoising_window_Q}, which we developed for the analysis of GW signals, can be applied to many other fields like music, medicine, geophysics, engineering and in general in all those fields which require the analysis of noisy, transient signals.

\section{Code availability}

A first implementation of the wavelet Q-- and Qp--transforms in Python, following the guidelines provided by the GWpy module \texttt{qtransform.py} \cite{macleod2021gwpy}, is freely available at the link \url{https://zenodo.org/doi/10.5281/zenodo.10649072}, including as an example the analysis of GW150914.

\section*{Acknowledgements}

We would like to thank the LIGO--Virgo groups of Trieste, Trento, Roma and Zürich for useful discussions.
This research has made use of data or software obtained from the Gravitational Wave Open Science Center (\url{http://gwosc.org}), a service of LIGO Laboratory, the LIGO Scientific Collaboration, the Virgo Collaboration, and KAGRA. LIGO Laboratory and Advanced LIGO are funded by the United States National Science Foundation (NSF) as well as the Science and Technology Facilities Council (STFC) of the United Kingdom, the Max--Planck--Society (MPS), and the State of Niedersachsen/Germany for support of the construction of Advanced LIGO and construction and operation of the GEO600 detector. Additional support for Advanced LIGO was provided by the Australian Research Council. Virgo is funded, through the European Gravitational Observatory (EGO), by the French Centre National de Recherche Scientifique (CNRS), the Italian Istituto Nazionale di Fisica Nucleare (INFN) and the Dutch Nikhef, with contributions by institutions from Belgium, Germany, Greece, Hungary, Ireland, Japan, Monaco, Poland, Portugal, Spain. KAGRA is supported by Ministry of Education, Culture, Sports, Science and Technology (MEXT), Japan Society for the Promotion of Science (JSPS) in Japan; National Research Foundation (NRF) and Ministry of Science and ICT (MSIT) in Korea; Academia Sinica (AS) and National Science and Technology Council (NSTC) in Taiwan. EM acknowledges support by ICSC – Centro Nazionale di Ricerca in High Performance Computing, Big Data and Quantum Computing, funded by European Union – NextGenerationEU.

\bibliographystyle{abbrv}
\bibliography{bibliography.bib}

\end{document}